\documentclass[twocolumn,showpacs,amsmath,amssymb,nofootinbib,amsfonts,showkeys]{revtex4}

\RequirePackage{graphicx}
\RequirePackage{amsmath}

\begin{document}

\title{Constraints on the interaction strength in the model of interacting dynamical dark energy with linear and non-linear interacting terms}

\author{R. G. Neomenko}
\email{roman.neomenko@lnu.edu.ua}

\affiliation{Astronomical Observatory of Ivan Franko National University of Lviv, Kyryla i Mefodiya Street 8, Lviv, 79005, Ukraine}

\date{\today}

\begin{abstract}
In this work the observational constraints on interaction coupling parameter between dynamical dark energy and cold dark matter were obtained using CMB, BAO and SN Ia data. The dark energy in considered models is dynamical and evolution of its equation of state parameter depends on dark coupling and internal properties of the dark energy. Such model is believed to be more physically consistent than models of interacting dark energy considered in previous works. Constraints were made for three types of interaction. The first two are the types which are often considered in other works on interacting dark energy. The third type has the non-linear dependence on densities of dark components and is studied for the first time. Observational constraints on Hubble constant $H_{0}$ for the first two models are in strong disagreement with so called local measurements of $H_{0}$. And the third model is in better agreement with local measurements than $\Lambda$CDM model. Also for the last non-linear model existence of non-zero interaction was found at greater than $1\sigma$ significance level.
\end{abstract}



\maketitle

\section{Introduction}
\label{sec:intro}
Interacting dark energy (IDE) is an extension of cosmological model which aim to explain the accelerated expansion of universe \cite{Riess1998,Perlmutter1999}. In this model some form of new interaction is present between dark energy (DE) which causes this acceleration and another component the dark matter (DM) in addition to the four known fundamental interactions \cite{Amendola2000,Zimdahl2001}. The existence of these dark components follows from their gravitational impact on visible matter and radiation as it didn't interact through other three fundamental forces. As the result the presence of such DE-DM interaction can be concluded if it make significant impact through gravitational interaction to make imprint on cosmic microwave background and other astrophysical data. This fact will be the possible indication that DE and DM have the quantum-field nature. In the most well studied IDE model the DE equation of state (EoS) parameter does not vary in time and the DE-DM interaction is proportional to the energy densities of DE, DM or to the sum of both and is generated by the expansion rate of universe \cite{Jackson2009,Gavela2009,DiValentino2017,Amendola2007,Chimento2003}. The constraints on parameters of such models using data on cosmic microwave background, baryon acoustic oscillations and type Ia supernova give the non-zero energy transfer between DE and DM with a low confidence level or interaction is absent at all. \cite{DiValentino2020,Pan2020}. Also the constraints on DE-DM interaction parameter proportional to the density of DM or to the density sum of dark components were made for phantom DE model only \cite{Costa2017,An2018}. For quintessence model such analysis is impossible due to non-adiabatic instabilities of cosmological perturbations in the radiation-dominated epoch of universe for these IDE models \cite{Valiviita2008}. But observational constraints for such models is possible when DE EoS parameter varies in time and its evolution can be tuned in such way that non-adiabatic instabilities will not arise. Hence in this work the Markov Chain Monte-Carlo constraints on parameters of dynamical quintessence IDE with these interaction forms were done for the first time. The model of quintessence IDE EoS parameter evolution that was used here was proposed in works \cite{Neomenko2016,Neomenko2017}, which is more physically consistent, than the well-known linear model for EoS evolution $w(a)=w_{0}+w_{1}(1-a)$ \cite{Neomenko2024}. The second part of this work is dedicated to the analysis of another type of DE-DM interaction (also for the first time), which is not depended on expansion rate of universe and has the form of Coulomb-type interaction function (e. g. the energy-momentum exchange rate between dark components is proportional to the product of densities of this components). Such interaction form is physically well motivated as it doesn't vanishes when universe is not expanding and its form often occurs among other interactions in nature also.

In the chapter I of this work the brief introduction into the models of dynamical IDE were analyzed is done. In the chapter II the observational data and method of statistical constraints which were used are described. And in the chapter III the impact of DE-DM interaction on the formation of high-scale structure of universe and the results of parameters' observational constraints of considered models are shown and discussed.

\section{I. Models of dynamical IDE}
\label{sec:ide}
The description of each component of universe is done in the perfect fluid approximation with the following stress-energy tensor:
\begin{equation}\label{tik}
T_{i}^{k}=(\rho+p)u_{i}u^{k}+p\delta_{i}^{k} \,.
\end{equation}
Universe is considered to be homogeneous and isotropic, which is described by Friedman-Lema\^{\i}tre-Robertson-Walker (FLRW) metric with zero spatial curvature in relation to which the small perturbations of metric (perturbations are given in synchronous gauge):
\begin{equation}\label{eq:ds2}
ds^{2}=a^{2}(\eta)[-d\eta^{2}+(\delta_{\alpha\beta}+h_{\alpha\beta})dx^{\alpha}dx^{\beta}]\,,
\end{equation}
where $a$ denotes a scale factor, $\eta$ is conformal time and $h_{\alpha\beta}$ is perturbation of metric tensor. For each component the general-covariant equation of stress-energy tensor conservation is true except for DE and DM, which in the result of non-gravitational interaction between them modifies into the following form:
\begin{subequations}\label{eq:tq}
\begin{gather}
\label{eq:tq:1}
T_{(de)i;k}^{k} = J_{(de)i} \,,
\\
\label{eq:tq:2}
T_{(c)i;k}^{k} = J_{(c)i} \,.
\end{gather}
\end{subequations}
Here ";" denotes the general-covariant derivative and $J_{i}$ is the 4-vector of energy-momentum exchange between DE and DM or in other words it describes the DE-DM interaction. The demand of conservation of energy and momentum of total DE and DM fluid implies that $J_{(c)i}=-J_{(de)i}=J_{i}$.

To solve the system of equations \eqref{eq:tq} along with Einstein's gravitational field equations the 4-vector $J_{i}$ must be given as function of variables which describe the state of DE and DM. In the most works on IDE this interaction is taken in the form which in FLRW universe is proportional to Hubble parameter $H$ and some function of dark components' densities $\bar{\rho}_{de}$, $\bar{\rho}_{c}$. In the cases considered in this work $\bar{J}_{0}$ is taken in the following forms \cite{Amendola2007,Chimento2003}:
\begin{subequations}\label{eq:fr}
\begin{gather}
\label{eq:fr:1}
\bar{J}_{0}=3\beta aH\bar{\rho}_{c} \,, \\
\label{eq:fr:2}
\bar{J}_{0}=3\beta aH(\bar{\rho}_{de}+\bar{\rho}_{c}) \,.
\end{gather}
\end{subequations}
Here $\beta$ is the interaction parameter and when it goes to zero the DE-DM interaction disappears. When the consideration of these interaction forms is extended on the small linear cosmological perturbations in the relation FLRW universe, then, as was mentioned above, the problem of instabilities of these perturbations in the radiation-dominated epoch occurs \cite{Valiviita2008}. To avoid this problem DE EoS parameter must be allowed to evolve with universe's expansion. In this study the model of IDE is considered which in the evolution of this EoS parameter is given by DE-DM interaction parameter and DE adiabatic sound speed. Consequently equations \eqref{eq:tq} with additional equation for DE EoS parameter evolution in FLRW universe take the following form:
\begin{subequations}\label{eq:rdedm}
\begin{gather}
\label{eq:rdedm:1}
\dot{\bar{\rho}}_{de}+3aH(1+w)\bar{\rho}_{de}=-\bar{J}_{0} \,,
\\
\label{eq:rdedm:2}
\dot{\bar{\rho}}_{c}+3aH\bar{\rho}_{c}=\bar{J}_{0} \,,
\\
\label{eq:rdedm:3}
\dot{w}=3aH(1+w)(w-c_{a}^{2})+\frac{\bar{J}_{0}}{\bar{\rho}_{de}}(w-c_{a}^{2}) \,.
\end{gather}
\end{subequations}
Here dot over quantity is the derivative on conformal time $\eta$, $w$ is the DE EoS parameter and $c_{a}^{2}=\dot{\bar{p}}_{de}/\dot{\bar{\rho}}_{de}$ is the square of DE adiabatic sound speed ($\bar{p}_{de}$ is the DE pressure). The solutions of these equations were obtained in works \cite{Neomenko2016,Neomenko2017}. To extend our models to the case of small perturbations in relation to the background universe the general-covariant form of DE-DM interactions must be given at first. In this study the following form is used which was proposed in \cite{Neomenko2020,Neomenko2021}:
\begin{subequations}\label{eq:jcs}
\begin{gather}
\label{eq:jcs:1}
J_{i}=\beta\rho_{c}u^{k}_{;k}u_{i}^{(c)} \,, \\
\label{eq:jcs:2}
J_{i}=\beta(\rho_{de}+\rho_{c})u^{k}_{;k}u_{i}^{(c)} \,,
\end{gather}
\end{subequations}
where $u_{i}^{(c)}$ is a four-vector of DM velocity and $u^{k}$ is a velocity four-vector of all components' center of mass.

Beside two forms of DE-DM interactions \eqref{eq:jcs}, in this study is considered another form of $J_{i}$ which is not generated by the expansion rate of universe. In other words it is not proportional to Hubble parameter $H$ in FLRW universe as the previous two types. Also such interaction is proportional to the product of DE and DM densities. So its general-covariant form is as follows:
\begin{equation}\label{jpr}
J_{i}=3\beta H_{0}\frac{\rho_{de}\rho_{c}}{\rho_{de}+\rho_{c}}u_{i}^{(c)} \,.
\end{equation}
The presence of Hubble constant $H_{0}$ in this interaction form is for normalization of interaction parameter $\beta$ only. The interaction \eqref{jpr} is motivated by those that frequently occur in different areas of physics such as Coulomb electrostatic interaction, Newtonian gravitational interaction and etc.. Such interaction is being studied for the first time.

The resulting equations for the evolution of cosmological perturbations for DE and DM with interaction \eqref{eq:jcs:1} in synchronous gauge comoving to DM are as follows:
\begin{subequations}\label{eq:prtbc}
\begin{gather}
\dot{\delta}_{de}=-3aH(c_{s}^{2}-w)\delta_{de}-(1+w)\frac{\dot{h}}{2}-
\nonumber\\
-(1+w)[k^{2}+9a^{2}H^{2}(c_{s}^{2}-c_{a}^{2})]\frac{\theta_{de}}{k^{2}}-
\nonumber\\
-\beta\frac{\bar{\rho}_{c}}{\bar{\rho}_{de}}\biggr[3aH(\delta_{c}-\delta_{de})+
\nonumber\\
+\frac{\dot{h}}{2}+\theta+9a^{2}H^{2}(c_{s}^{2}-c_{a}^{2})\frac{\theta_{de}}{k^{2}}\biggl]\,,\label{eq:prtbc:1}
\\
\dot{\theta}_{de}=-aH(1-3c_{s}^{2})\theta_{de}+\frac{c_{s}^{2}k^{2}}{1+w}\delta_{de}+
\nonumber\\
+3aH\frac{\beta}{1+w}\frac{\bar{\rho}_{c}}{\bar{\rho}_{de}}(1+c_{s}^{2})\theta_{de}\,,
\label{eq:prtbc:2}
\\
\dot{\delta}_{c}=-\frac{\dot{h}}{2}+\beta\biggr[\frac{\dot{h}}{2}+\theta\biggl]\,,
\label{eq:prtbc:3}
\end{gather}
\end{subequations}
where $\theta_{N}\equiv i(\overrightarrow{k},\overrightarrow{v}_{N})$,  $c_{s}^{2}$ is a comoving effective DE sound speed which in this work is taken as $c_{s}^{2}=1$ and
$$\theta=\frac{\sum_{N}(\bar{\rho}_{N}+\bar{p}_{N})\theta_{N}}{\sum_{N}(\bar{\rho}_{N}+\bar{p}_{N})}\,,$$
where $N$ is a number of universe's each component.

For the interaction \eqref{eq:jcs:2} we have such equations:
\begin{subequations}\label{eq:prtbs}
\begin{gather}
\dot{\delta}_{de}=-3aH(c_{s}^{2}-w)\delta_{de}-(1+w)\frac{\dot{h}}{2}-
\nonumber\\
-(1+w)[k^{2}+9a^{2}H^{2}(c_{s}^{2}-c_{a}^{2})]\frac{\theta_{de}}{k^{2}}-
\nonumber\\
-\beta\frac{\bar{\rho}_{de}+\bar{\rho}_{c}}{\bar{\rho}_{de}}\biggr[3aH\biggl(\frac{\bar{\rho}_{de}}{\bar{\rho}_{de}+\bar{\rho}_{c}}\delta_{de}+
\nonumber\\
+\frac{\bar{\rho}_{c}}{\bar{\rho}_{de}+\bar{\rho}_{c}}\delta_{c}-\delta_{de}\biggr)+
\nonumber\\
+\frac{\dot{h}}{2}+\theta+9a^{2}H^{2}(c_{s}^{2}-c_{a}^{2})\frac{\theta_{de}}{k^{2}}\biggl]\,,\label{eq:prtbs:1}
\\
\dot{\theta}_{de}=-aH(1-3c_{s}^{2})\theta_{de}+\frac{c_{s}^{2}k^{2}}{1+w}\delta_{de}+
\nonumber\\
+3aH\frac{\beta}{1+w}\frac{\bar{\rho}_{de}+\bar{\rho}_{c}}{\bar{\rho}_{de}}(1+c_{s}^{2})\theta_{de}\,,
\label{eq:prtbs:2}
\\
\dot{\delta}_{c}=-\frac{\dot{h}}{2}+\beta\frac{\bar{\rho}_{de}+\bar{\rho}_{c}}{\bar{\rho}_{c}}\biggr[3aH\biggl(\frac{\bar{\rho}_{de}}{\bar{\rho}_{de}+\bar{\rho}_{c}}\delta_{de}+
\nonumber\\
+\frac{\bar{\rho}_{c}}{\bar{\rho}_{de}+\bar{\rho}_{c}}\delta_{c}-\delta_{c}\biggr)+\frac{\dot{h}}{2}+\theta\biggl]\,.
\label{eq:prtbs:3}
\end{gather}
\end{subequations}

And for the interaction \eqref{jpr}:
\begin{subequations}\label{eq:prtbp}
\begin{gather}
\dot{\delta}_{de}=-3aH(c_{s}^{2}-w)\delta_{de}-(1+w)\frac{\dot{h}}{2}-
\nonumber\\
-(1+w)[k^{2}+9a^{2}H^{2}(c_{s}^{2}-c_{a}^{2})]\frac{\theta_{de}}{k^{2}}-
\nonumber\\
-3\beta aH_{0}\frac{\bar{\rho}_{c}}{\bar{\rho}_{de}+\bar{\rho}_{c}}\biggr[\delta_{c}+3aH(c_{s}^{2}-c_{a}^{2})\frac{\theta_{de}}{k^{2}}-
\nonumber\\
-\frac{\bar{\rho}_{de}}{\bar{\rho}_{de}+\bar{\rho}_{c}}\delta_{de}-\frac{\bar{\rho}_{c}}{\bar{\rho}_{de}+\bar{\rho}_{c}}\delta_{c}\biggl]\,,\label{eq:prtbp:1}
\\
\dot{\theta}_{de}=-aH(1-3c_{s}^{2})\theta_{de}+\frac{c_{s}^{2}k^{2}}{1+w}\delta_{de}+
\nonumber\\
+\frac{3aH_{0}\beta}{1+w}\frac{\bar{\rho}_{c}}{\bar{\rho}_{de}+\bar{\rho}_{c}}(1+c_{s}^{2})\theta_{de}\,,
\label{eq:prtbp:2}
\\
\dot{\delta}_{c}=-\frac{\dot{h}}{2}+3\beta aH_{0}\frac{\bar{\rho}_{de}}{\bar{\rho}_{de}+\bar{\rho}_{c}}\biggr[\delta_{de}-\nonumber\\
-\frac{\bar{\rho}_{de}}{\bar{\rho}_{de}+\bar{\rho}_{c}}\delta_{de}-\frac{\bar{\rho}_{c}}{\bar{\rho}_{de}+\bar{\rho}_{c}}\delta_{c}\biggl]\,.
\label{eq:prtbp:3}
\end{gather}
\end{subequations}

To make numerical integration of this system of equations the initial conditions for the background system \eqref{eq:rdedm} and for perturbed system \eqref{eq:prtbc}, \eqref{eq:prtbs}, \eqref{eq:prtbp} must be set up. The background initial conditions are given at present epoch at $a_{0}=1$ and the perturbation initial conditions are given at early epoch of electro-magnetic radiation dominance.

Initial conditions for perturbation equations are taken as their solutions at radiation-dominated epoch when the perturbations not enter yet into the Hubble horizon. These solutions satisfy the following condition for the arbitrary two components $x$ and $y$
\begin{equation}\label{sdmr}
 S_{x,y}=aH\left(\frac{\delta_{x}}{(\dot{\bar{\rho}}_{x}/\bar{\rho}_{x})}-\frac{\delta_{y}}{(\dot{\bar{\rho}}_{y}/\bar{\rho}_{y})}\right)=0\,,
\end{equation}
and as the result the fluids are adiabatic. When DE does not interact with DM the small deviations from adiabatic perturbations are damped and as the result these perturbations stay stable until they enter into the Hubble horizon. But when DE-DM interaction of form \eqref{eq:jcs:1} or \eqref{eq:jcs:2} is present and DE is quintessential this adiabatic mode could become unstable if DE EoS parameter $w$ is close $-1$ \cite{Valiviita2008}. To avoid this problem the stability analysis of adiabatic solutions of perturbation equations was made. From this the ranges of values for interaction parameter and DE adiabatic sound speed $c_{a}^{2}$ for which adiabatic mode is stable were derived \cite{Neomenko2020,Neomenko2021} for each of the interactions \eqref{eq:jcs:1}, \eqref{eq:jcs:2}. It must be noted that for interaction of type \eqref{jpr} the adiabatic mode, as it follows from analysis, at the early epoch is always stable. Hence for all three types of DE-DM interaction there can be used the standard adiabatic initial conditions without interaction even if they differ by a small value from the true initial conditions with non-zero interaction, because as was mention above small deviations in the true initial conditions disappear.

\section{II. Observational data and statistical method}
\label{sec:odsm}

To make the constraints on parameters of IDE models \eqref{eq:fr:1} (it will be called Model I), \eqref{eq:fr:2} (Model II) and \eqref{jpr} (Model III) the following observational data were used:

{\bf Cosmic Microwave Background (CMB) anisotropies:} the dataset consisting of high-$l$ TT, EE, TE power spectra and low-$l$ TT, EE power spectra of Planck collaboration (2018 data release) \cite{CMB2018}; This dataset is supplemented by additional data on CMB weak gravitational lensing of same collaboration (2018 data release) \cite{Lensing2018};

{\bf Baryon Acoustic Oscillations (BAO):} the 6dF Galaxy Survey \cite{6dF} consisting of one data point at effective redshift $z_{eff}=0.106$, SDSS DR7 Main Galaxy Sample \cite{DR7} of data point at $z_{eff}=0.15$ and SDSS-III Baryon Oscillation Spectroscopic Survey, DR12 \cite{DR12} consisting of three data points at $z_{eff}=0.38,\, 0.51,\, 0.61$;

{\bf Type Ia supernova (SN Ia):} Pantheon dataset consisting of data on 1048 type Ia supernova \cite{Pantheon}.

To confront the Models I, II and III with these observational data the corresponding observable quantities must be calculated. For this purpose was modified the code IDECAMB \cite{Li2023} which is the modification of program package CAMB \cite{CAMB} and is specially designed for considering IDE models. In this program to calculate the cosmological perturbations' evolution the Parameterized Post Friedman (PPF) method adapted for IDE models was used \cite{Li2014}. To be suitable for Models I and II it must take into account the local Hubble parameter perturbations described by perturb part of $u^{k}_{;k}$ in expressions \eqref{eq:jcs}. It was done by modifying the expressions (3.14) and (3.15) given in work \cite{Li2023}:
\begin{subequations}\label{eq:int}
\begin{gather}
\label{eq:int:1}
\Delta Q=C_{1}\delta_{de}+C_{2}\delta_{c}+Q\biggl(\frac{kV}{3aH}+\frac{\zeta^{\prime}}{aH}-\xi\biggr) \,, \\
\label{eq:int:2}
f_{k}=Q(\theta_{c}-\theta) \,,
\end{gather}
\end{subequations}
where $\bar{J}_{0}=-aQ$ and $\zeta^{\prime}$ is given by expression (4.9) in \cite{Li2023}. For Model I $Q=-3\beta H\bar{\rho}_{c}$, $C_{1}=0$, $C_{2}=Q$ and for Model II $Q=-3\beta H(\bar{\rho}_{de}+\bar{\rho}_{c})$,
\begin{gather}
\label{eq:qs12:2}
C_{1}=\frac{\bar{\rho}_{de}}{\bar{\rho}_{de}+\bar{\rho}_{c}}Q \,, \quad C_{2}=\frac{\bar{\rho}_{c}}{\bar{\rho}_{de}+\bar{\rho}_{c}}Q \,. \nonumber
\end{gather}
For Model III the expression \eqref{eq:int:1} takes the following form:
\begin{equation}\label{int3}
\Delta Q=C_{1}\delta_{de}+C_{2}\delta_{c} \,,
\end{equation}
where
\begin{subequations}\label{eq:qc12}
\begin{gather}
\label{eq:qc12:1}
Q=-3\beta H_{0}\frac{\bar{\rho}_{de}\bar{\rho}_{c}}{\bar{\rho}_{de}+\bar{\rho}_{c}} \,, \nonumber\\
\label{eq:qc12:2}
C_{1}=\frac{\bar{\rho}_{c}}{\bar{\rho}_{de}+\bar{\rho}_{c}}Q \,, \quad C_{2}=\frac{\bar{\rho}_{de}}{\bar{\rho}_{de}+\bar{\rho}_{c}}Q \,. \nonumber
\end{gather}
\end{subequations}

\begin{table}
\begin{center}
\caption{Priors of independent parameters for each IDE model.}
\begin{tabular} {|l|l|l|l|l|}
\hline
\hline
   \noalign{\smallskip}
Parameter& $\Lambda$CDM &Model I&Model II&Model III  \\
 \noalign{\smallskip}
\hline
   \noalign{\smallskip}
$\Omega_{b}h^{2}$ &[0.005, 0.1] & [0.005, 0.1] & [0.005, 0.1] & [0.005, 0.1] \\
$\Omega_{c}h^{2}$ &[0.001, 0.99] & [0.001, 0.99] & [0.001, 0.99] & [0.001, 0.99] \\
$100\theta_{MC}$ &[0.5, 10] & -- & -- & [0.5, 10] \\
$H_{0}$ & -- & [40, 100] & [40, 100] & -- \\
$\tau$ &[0.01, 0.8] & [0.01, 0.8] & [0.01, 0.8] & [0.01, 0.8] \\
$\log(10^{10}A_{s})$ &[1.61, 3.91] & [1.61, 3.91] & [1.61, 3.91] & [1.61, 3.91] \\
$n_{s}$ &[0.8, 1.2] & [0.8, 1.2] & [0.8, 1.2] & [0.8, 1.2] \\
$w_{0}$ & -- & [-1, -0.333] & [-1, -0.333] & [-3, -0.333] \\
$c_{a}^{2}$ & -- & [-1, 0] & [-0.533890, 0] & [-3, -1] \\
$\beta$ & -- & [0, 0.08] & [0, 0.5] & [-1.5, 1.5] \\
    \noalign{\smallskip}
  \hline
  \hline
\end{tabular}
\label{tab:1}
\end{center}
\end{table}

Constraints on interaction parameters and other parameters of IDE models were obtained using Markov Chain Monte-Carlo method with modifying of CosmoMC program package \cite{CosmoMC} for this purpose. There were ran 12 Monte-Carlo chains for each of the studied IDE models with a convergence condition (using the Gelman-Rubin parameter) to be $R-1<0.01$. The priors for independent parameters which describe the pressure of DE $w_{0}$ and $c_{a}^{2}$ were taken in quintessence range of values and for interaction parameter $\beta$ they were taken in positive range of values (case when energy flows from DE to DM) for Model I and Model II. Also the additional priors for these models were taken derived from positivity of energy density of dark components conditions \cite{Neomenko2016,Neomenko2021} and stability of early cosmological perturbations conditions \cite{Neomenko2020,Neomenko2021}. For Model III priors for $c_{a}^{2}$ were taken in phantom range, for $w_{0}$ they were taken in quintessence and phantom ranges and $\beta$ is bounded by negative lower value and positive upper value. For Models I, II the $H_{0}$-parametrization was used and for Model III the $100\theta_{MC}$-parametrization was used. Beside this there were ran 12 Monte-Carlo chains for $\Lambda$CDM model with the same observational data to compare its constraints with the results for IDE models. Priors for $\Lambda$CDM and all three interaction models are given in Table \ref{tab:1}.

\begin{figure}
\includegraphics[width=0.5\textwidth]{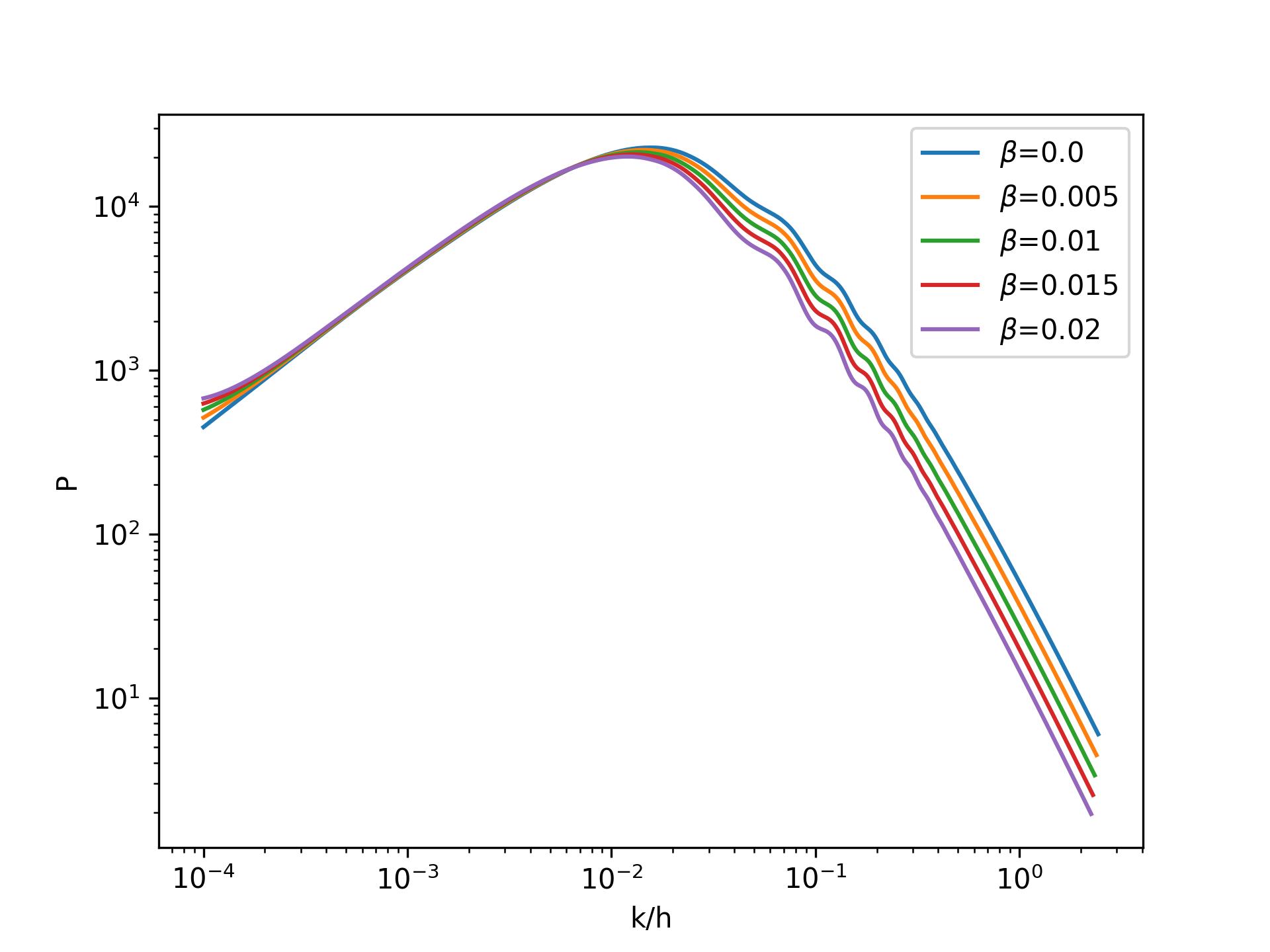}
\includegraphics[width=0.5\textwidth]{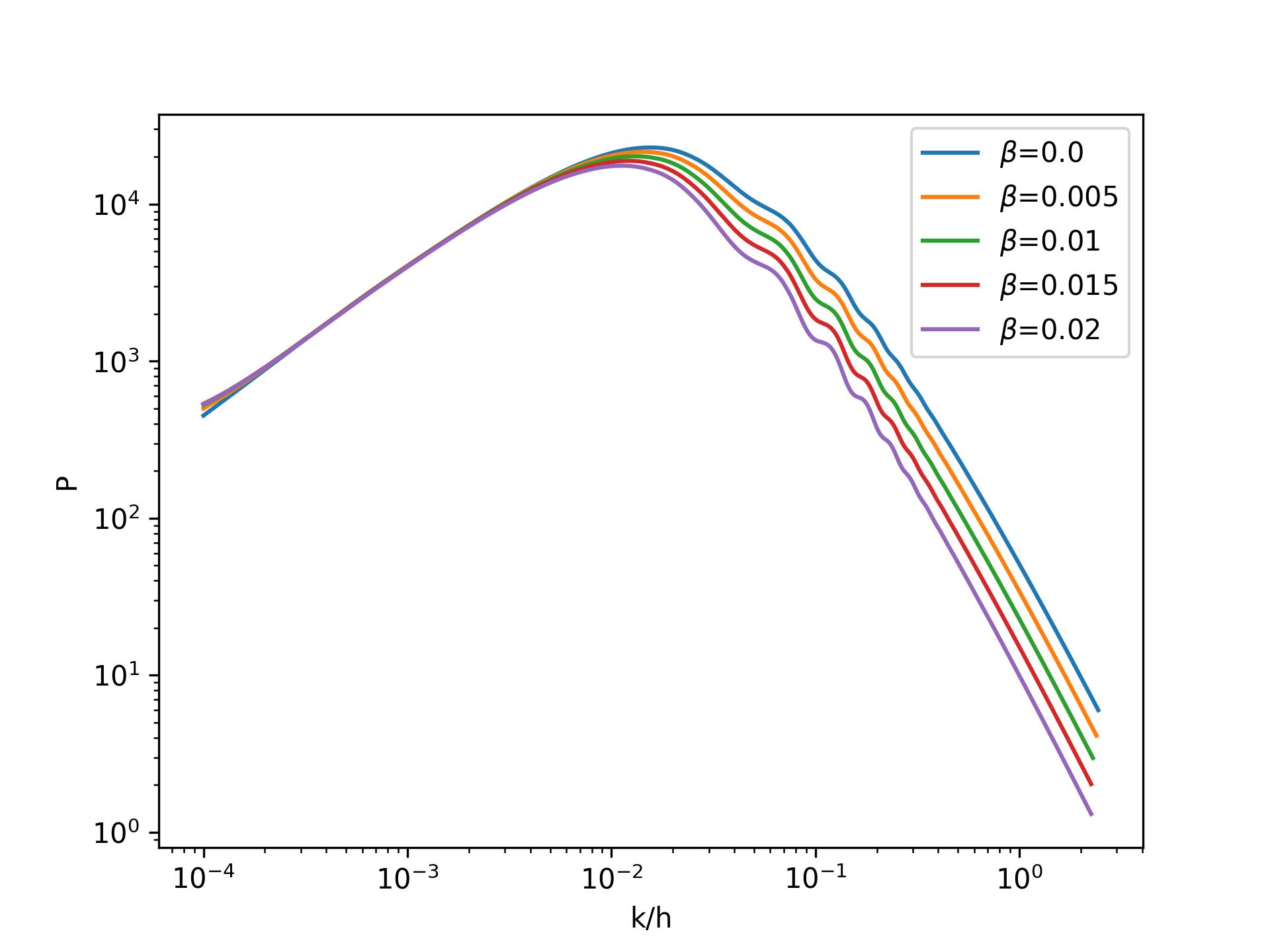}
\caption{Dependence of the matter power spectrum at redshift $z=0$ on the interaction parameter $\beta$ for Model I in the upper panel and for Model II in the lower panel. The independent model parameters that were used are as follows: $\Omega_{b}h^{2}=0.0226$, $\Omega_{c}h^{2}=0.112$, $H_{0}=68.2$, $\Omega_{K}=0$, $A_{s}=2.1\times10^{-9}$, $n_{s}=0.96$, $\tau=0.09$, $c_{s}^{2}=1$, $w_{0}=-0.9$, $c_{a}^{2}=-0.5$.}
\label{fig:1}
\end{figure}

\begin{figure}
\includegraphics[width=0.5\textwidth]{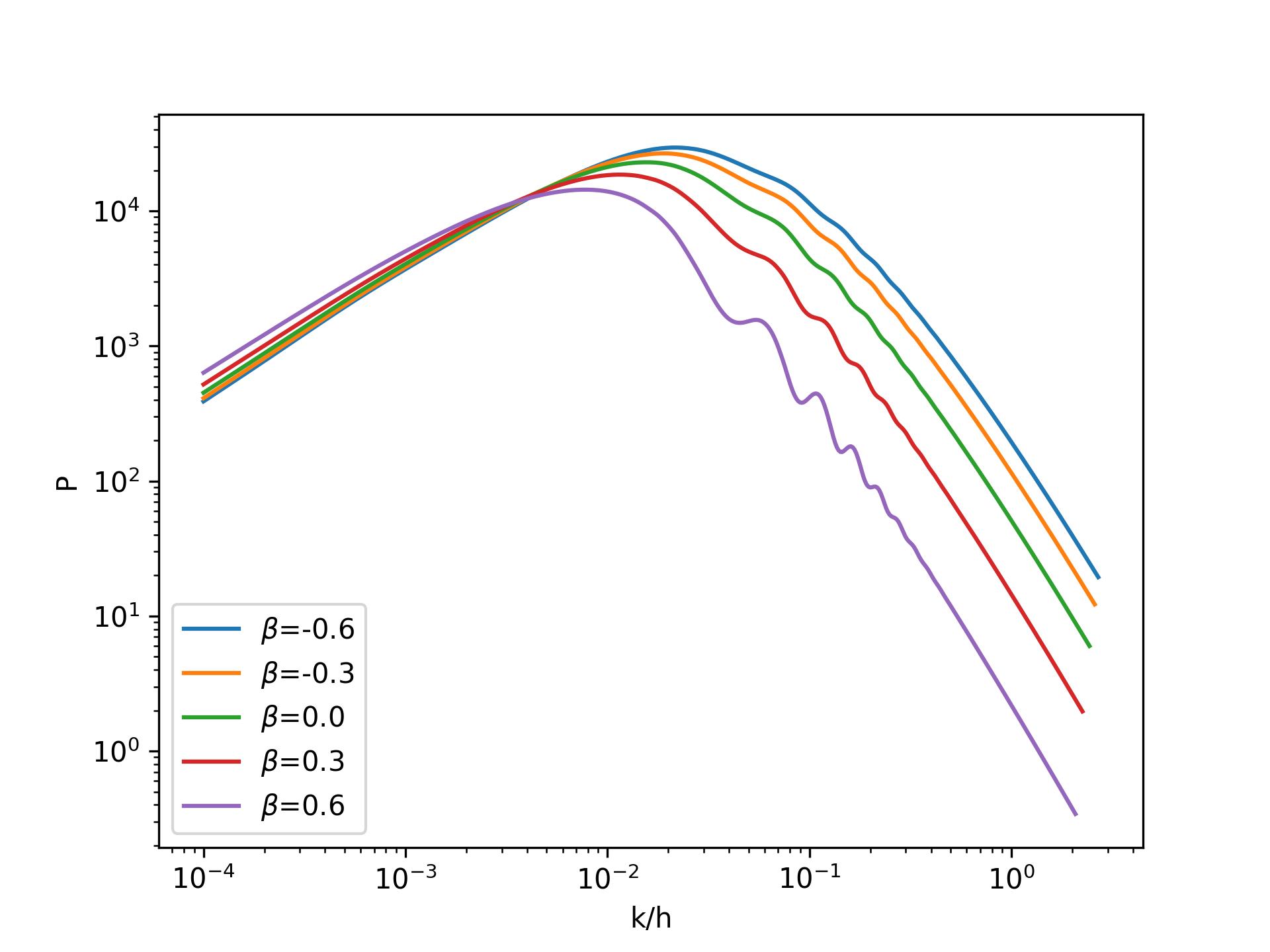}
\includegraphics[width=0.5\textwidth]{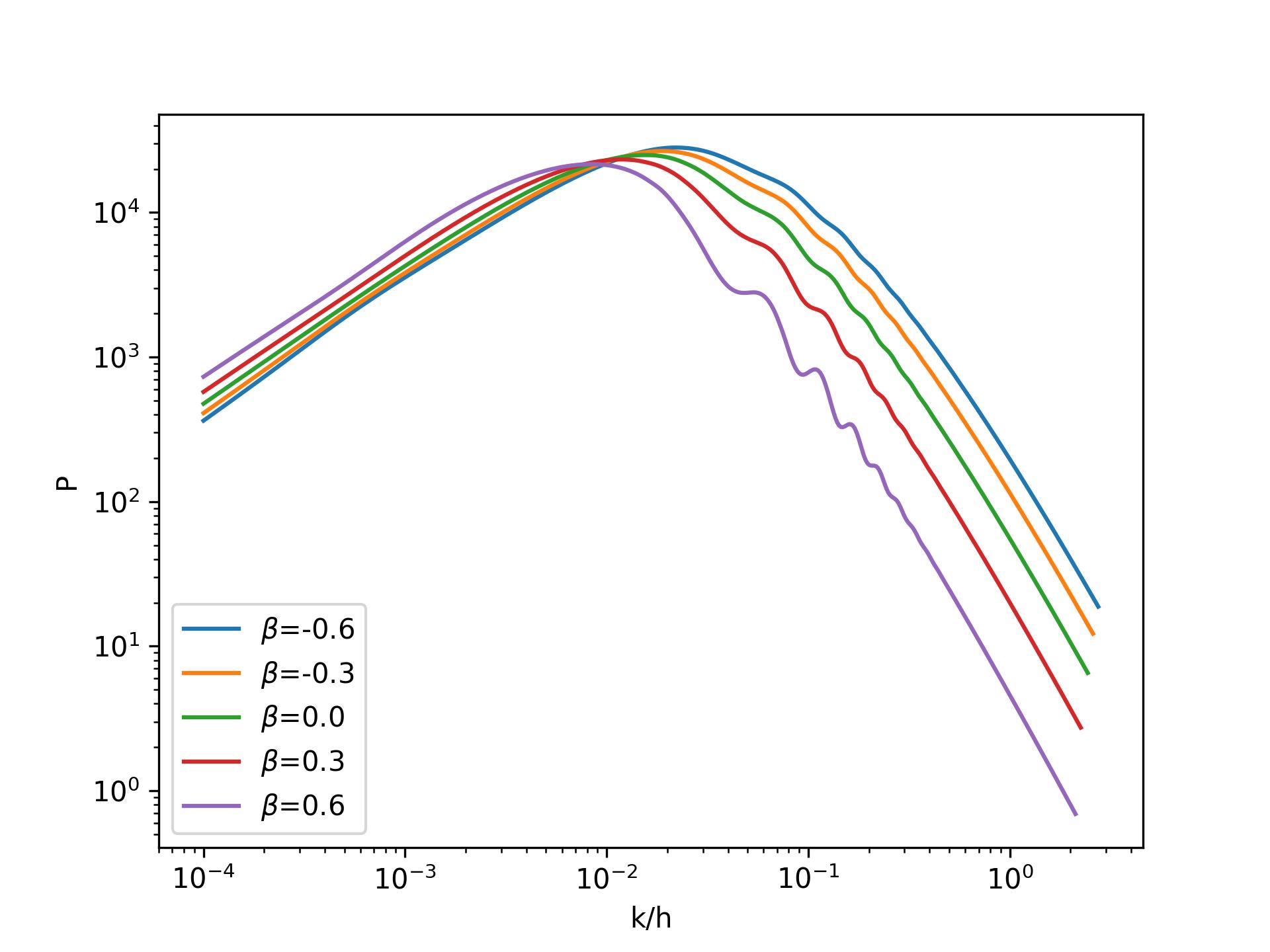}
\caption{Dependence of the matter power spectrum at redshift $z=0$ on the interaction parameter $\beta$ for Model III with $c_{a}^{2}=-0.5$ in the upper panel and $c_{a}^{2}=-1.2$ in the lower panel. The independent model parameters that were used are the same as in Fig. \ref{fig:1}.}
\label{fig:2}
\end{figure}

\section{III. Results}
\label{sec:res}
At first the dependence of high scale structure of universe on DE-DM interaction coupling was studied for Model I, Model II and Model III. In Fig. \ref{fig:1} and Fig. \ref{fig:2} the modification of matter power spectrum at redshift $z=0$ by the value of interaction parameter $\beta$ is shown for all of three models. For Models I and II the modifications are similar with the suppression of structure formation at small scales and with some larger inhomogeneities at very high scales compared to non-interacting case (in these figures for Models I, II the interaction parameter $\beta$ is bounded to positive values only, the same as in priors in MCMC simulations). For Model III the distribution of matter in universe is more inhomogeneous at high scales and sufficiently more homogeneous at small scales when $\beta$ is positive. When we have negatively-valued $\beta$ (it's corresponding to the case when energy flows from DM to DE) the impact of DE-DM interaction is exactly opposite -- at high scales matter is distributed slightly less homogeneous and at small scales the matter structure growth is larger.

The observational constraints on parameters of Model I, Model II and Model III obtained from MCMC simulation at $68$\% CL are given in Table \ref{tab:2}.

As we can see for the quintessence IDE of Model I and Model II due to presence of DE-DM interaction the relative part of DE component is much lower and DM much higher compared to $\Lambda$CDM model. As the result the Hubble constant $H_{0}$ is much lower than the value obtained in work \cite{Riess2022}. So, such models only worsen the so called Hubble tension, which is one of the major problems in modern cosmology. Also for both of these models only the upper positive bounds on interaction parameter $\beta$ were obtained. The constraints on DE EoS parameter at present time $w_{0}$ and EoS parameter evolution which mostly determined by DE squared adiabatic sound speed $c_{a}^{2}$ are strongly prefer the dynamical nature of quintessence DE. In general the constraints on $\beta$ using CMB, BAO and SN Ia data described in chapter \ref{sec:odsm} do not allow to determine whether DE-DM interaction of Model I and Model II exists.

The constraints for Model III on interaction parameter $\beta$ give the existence of its non-zero positive value on $>1\sigma$ significance level. Also constraints on EoS parameter prefer of DE having the quintessential nature in epochs closer to modern time and being phantom in the early epochs of universe. It means that DE energy density $\bar{\rho}_{de}$ begins to increase from some constant value after universe's expansion start and after approaching some maximum follows the gradual decrease of DE density till present epoch at $a=1$. Such model (but in non-interacting case) was studied in the work \cite{Novosyadlyj2013}. In this model there are higher proportion of DM and lower of DE components compare to $\Lambda$CDM model as in the previous two IDE models. The Hubble constant $H_{0}$ in Model III constraints has a bit higher value than in $\Lambda$CDM model which may indicate that this model maybe could give the resolution of Hubble tension if for the constraining its parameters the next generation BAO and SN Ia data will be used.

\section{Conclusions}
\label{sec:con}
In this work the cosmological models of interacting dynamical dark energy were studied in which non-gravitational interaction between dynamical dark energy and dark matter is present and is described by three different functions. The first two are well known in literature functions which are proportional to Hubble parameter and one is also proportional to the dark matter energy density (Model I) and another to the sum of energy densities of both dark components (Model II). The third one is not depended on expansion rate of universe and is proportional the product of energy densities of interacting components (Model III). Such interaction is studied for the first time and must be more physically realistic in comparison to previous two types of interaction and other types which are proportional to the Hubble parameter. By making Markov Chain Monte-Carlo constraints on parameters of these three models using the CMB, BAO and SN Ia data it was found that Models I, II are in large disagreement in determining the Hubble constant $H_{0}$ compare to the so called local measurement of $H_{0}$. Also it were determined only the upper bounds of interaction parameter for these models. In contrary Model III is in the more good agreement in determining $H_{0}$ with local measurements then $\Lambda$CDM model using the same observational data. Also the constraints give the non-zero positive value of interaction parameter (which corresponds to the energy flow from dark energy to dark matter) at $>1\sigma$ significance level for Model III. It is expected that using next-generation data on BAO and SN Ia along with current CMB data will give more tight constraints on the interaction in the dark sector.

\onecolumngrid
\begin{center}
\begin{table}
\caption{Constraints on model parameters at $68$\% CL.}
\begin{tabular} { l  c c c c}

 Parameter & $\Lambda$CDM & Model I & Model II & Model III\\
\hline
{\boldmath$\Omega_b h^2   $} & $0.02242\pm 0.00014        $ & $0.02282\pm 0.00015        $ & $0.02282\pm 0.00015        $ & $0.02239\pm 0.00014        $\\

{\boldmath$\Omega_c h^2   $} & $0.11932\pm 0.00092        $ & $0.1142\pm 0.0010          $ & $0.1142\pm 0.0011          $ & $0.151^{+0.027}_{-0.017}   $\\

{\boldmath$\tau           $} & $0.0573\pm 0.0074          $ & $0.083\pm 0.010            $ & $0.083\pm 0.010            $ & $0.0539\pm 0.0074          $\\

{\boldmath$w_{0}          $} &    ---                       & $-0.99424^{+0.00086}_{-0.0057}$ & $< -0.994                  $ & $-0.83^{+0.17}_{-0.29}     $\\

{\boldmath$c_{a}^{2}      $} &    ---                       & $-0.24553^{+0.00075}_{-0.0046}$ & $-0.24557^{+0.00076}_{-0.0046}$ & $-1.130^{+0.083}_{-0.074}  $\\

{\boldmath$\beta         $} &    ---                       & $< 9.43\cdot 10^{-5}       $ & $< 9.71\cdot 10^{-5}       $ & $0.27^{+0.23}_{-0.15}      $\\

{\boldmath${\rm{ln}}(10^{10} A_s)$} & $3.049\pm 0.014     $ & $3.092\pm 0.020            $ & $3.092\pm 0.020            $ & $3.042\pm 0.014            $\\

{\boldmath$n_s            $} & $0.9664\pm 0.0037          $ & $0.9805\pm 0.0041          $ & $0.9804\pm 0.0041          $ & $0.9658\pm 0.0040          $\\

$H_0                       $ & $67.66\pm 0.42             $ & $56.51\pm 0.25             $ & $56.51\pm 0.25             $ & $68.37\pm 0.83             $\\

$\Omega_{de}               $ & $0.6889\pm 0.0056          $ & $0.5690\pm 0.0066          $ & $0.5688\pm 0.0066          $ & $0.627^{+0.041}_{-0.055}   $\\

$\Omega_m                  $ & $0.3111\pm 0.0056          $ & $0.4310\pm 0.0066          $ & $0.4312\pm 0.0066          $ & $0.373^{+0.055}_{-0.041}   $\\

$\sigma_8                  $ & $0.8110\pm 0.0060          $ & $0.6843\pm 0.0067          $ & $0.6843\pm 0.0067          $ & $0.756^{+0.036}_{-0.055}   $\\

$S_8                       $ & $0.826\pm 0.011            $ & $0.820\pm 0.011            $ & $0.820\pm 0.011            $ & $0.839^{+0.012}_{-0.011}   $\\
\hline
\end{tabular}
\label{tab:2}
\end{table}
\end{center}

\end{document}